# Formation of calcium sulfate through the aggregation of sub-3 nanometre primary species

Tomasz M. Stawski[1,2], Alexander E.S. van Driessche[3,4,5], Mercedes Ossorio[3], Juan Diego Rodriguez-Blanco[6], Rogier Besselink[2] & Liane G. Benning[1,2]

The formation pathways of gypsum remain uncertain. Here, using truly *in situ* and fast time-resolved small-angle X-ray scattering, we quantify the four-stage solution-based nucleation and growth of gypsum ($CaSO_4 \cdot 2H_2O$), an important mineral phase on Earth and Mars. The reaction starts through the fast formation of well-defined, primary species of $<3$ nm in length (stage I), followed in stage II by their arrangement into domains. The variations in volume fractions and electron densities suggest that these fast forming primary species contain Ca–$SO_4$-cores that self-assemble in stage III into large aggregates. Within the aggregates these well-defined primary species start to grow (stage IV), and fully crystalize into gypsum through a structural rearrangement. Our results allow for a quantitative understanding of how natural calcium sulfate deposits may form on Earth and how a terrestrially unstable phase-like bassanite can persist at low-water activities currently dominating the surface of Mars.

[1] School of Earth and Environment, University of Leeds, Leeds LS2 9JT, UK. [2] German Research Centre for Geosciences, GFZ, 14473 Potsdam, Germany. [3] LEC, IACT, CSIC-UGR, E-18100 Armilla, Spain. [4] Structural Biology Brussels, VUB, 1050 Brussels, Belgium. [5] CNRS, ISTerre, F-38041 Grenoble, France. [6] Nano-Science Center; Department of Chemistry; University of Copenhagen, DK 2100 Copenhagen, Denmark. Correspondence and requests for materials should be addressed to T.M.S. (email: stawski@gfz-potsdam.de) or to A.E.S.v.D (email: alexander.van-driessche@ujf-grenoble.fr) or to L.G.B. (email: benning@gfz-potsdam.de).

Calcium sulfate has three crystalline phases with various hydration levels[1]: $CaSO_4 \cdot 2H_2O$ (gypsum), $CaSO_4 \cdot 0.5H_2O$ (bassanite) and $CaSO_4$ (anhydrite). Large natural (often evaporitic) deposits of gypsum and anhydrite exist on Earth[2], but at the same time bassanite is unstable at Earth surface conditions. Recently, significant amounts of bassanite and gypsum have been reported on Mars[3,4], yet why a thermodynamically unstable phase-like bassanite remains present on Mars is still debated[4]. The dihydrate gypsum is transformed to the hemihydrate bassanite only through an energy intensive dehydration[5]. Because tremendous quantities[6,7] of such 'synthetic' bassanite need to be produced for use in the construction industry as plaster of Paris (globally ~100 billion kg per year), ways to improve the energy efficiency of this production process are very much welcomed.

It is thus not surprising that a plethora of studies focused on elucidating the nucleation and growth mechanism of calcium sulfate polymorphs, for example, (refs 8–11; we use the terms polymorph and phase indistinctively and we refer to polymorphs *sensu lato*). Traditionally their formation from aqueous solutions was assumed to occur via a single-step process that is, following the classical nucleation paradigm. However, recent experimental observations contradict this classical picture and various possible intermediate phase(s), for example, amorphous calcium sulfate and/or nano-crystalline bassanite have been suggested as precursors to gypsum[12–15]. These intermediates were detected using primarily (cryo-) transmission electron microscopy (TEM) and/or infrared or Raman spectroscopies. Jones[16], based on infrared spectroscopy, suggested that gypsum formed from a long-lived disordered precursor phase. In the proposed crystallization mechanism, water molecules move from disordered positions to the ordered ones, in which water could be associated with the formation of crystalline gypsum and possibly bassanite. Wang *et al.*[12] showed by TEM imaging of time-resolved quenched samples that although the formation of gypsum was preceded by bassanite, an amorphous calcium sulfate phase was the first phase to nucleate from solution before bassanite. In our previous work[14], using high-resolution TEM analysis of time-resolved, cryo-quenched samples, we documented that the first and only phase that homogeneously precipitated from solution was nano-bassanite (5 to >10 nm) and not an amorphous phase. Subsequently these nano-bassanite particles self-assembled into elongated aggregates and eventually transformed into gypsum[14].

The primary reason for these ambiguities originates from the fact that it is extremely challenging to target the elusive early stages of the formation of a solid from aqueous solutions. The aforementioned recent studies mostly relied on high-resolution imaging and analyses of particles that were present in the pre-gypsum solid stages. Although such approaches offer in the best of cases near-atomic resolution, they also present clear disadvantages: imaging of particles is done *ex situ*, after quenching[12,14,15], under high vacuum and at high-beam energies and thus possible beam damage effects need to be considered. Hence, true *in situ* and time-resolved characterization of the formation reactions in solution could not be achieved in any of the above studies and this lead to the current uncertainty about the formation pathway(s) of gypsum.

To overcome this impasse, we performed synchrotron-based *in situ* and highly temporally resolved X-ray small- and wide-angle scattering experiments (SAXS/WAXS) in a way that we could follow all nucleation, growth and transformation reactions in solutions. Using this data we quantified each reaction step starting at the very earliest stages until a stable phase was fully developed. We show that instantly after reaching the desired supersaturation levels, well-defined, sub-3 nm in length species formed. These entities constitute the primary building blocks, which through aggregation, self-assembly and a structural rearrangement finally transform to gypsum. Based on these new insights we propose a non-classical four-stage pathway for calcium sulfate formation from homogenous solution.

## Results

**General analysis of the time-resolved scattering patterns**. We investigated the formation of solid $CaSO_4$ phases from various supersaturated solutions with initial concentrations between 50 and 150 mmol l$^{-1}$ and at temperatures between 12 and 40 °C (full details in Methods section). The analysis of the time-resolved SAXS patterns indicated that the scattering features for all experiments evolved in an equivalent manner, regardless of solution conditions. Thus, for simplicity, we discuss all observed changes in the structural information contained within the scattering patterns based on the 50 mmol l$^{-1}$ $CaSO_4$ experiments measured at 12 °C (for all other experiments see Supplementary Note 1, Supplementary Table 1, and Supplementary Fig. 1). At 12 °C a 50-mmol l$^{-1}$ solution is supersaturated with respect to gypsum and anhydrite, but undersaturated with respect to bassanite (Methods section, Table 1). The reaction kinetics at these conditions are relatively slow[14,17,18] and thus both the SAXS and WAXS patterns were collected at a time resolution of 30 s per frame and for ~4 h. This allowed us to capture all details of the reaction from the earliest stages to the final products. This experiment was used to develop and validate the model for the entire reaction pathway.

Figure 1 shows an overview of the scattering patterns for the first 5,400 s (1.5 h) of the reaction clearly indicating that over this time period the scattering intensity increased by several orders of magnitude. Although the experiment lasted for nearly 4 h (that is, 14,370 s), after ~5,400 s the overall SAXS signal remained relatively constant increasing by <5% during the final 2.5 h of the reaction. Based on this data set, four characteristic reaction stages could be distinguished (Fig. 1): stage I, 30–120 s: formation of small primary entities/scatterers as evidenced through the change in $I(q) \propto q^{-1}$ for $q > 1$ nm$^{-1}$ and $I(q) \propto q^0$ for $q < 1$ nm$^{-1}$; stage II, 150–390 s: development of a structure factor, manifested by the decrease of intensity at $q < 0.3$ nm$^{-1}$, indicating interactions between the previously formed primary scatterers; stage III, 420–1,500 s (and up to 5,400 s): formation and development of large scattering features evidenced through the increase in intensity at $q < 1.0$ nm$^{-1}$; this change followed a dependence of $I(q) \propto q^{-3 > a > -4}$ (where $a$ is the exponent); the overall intensity and shape of the curves at $q > 1.0$ nm$^{-1}$ corresponding to the scattering from the primary species remained relatively unchanged at this stage; stage IV, >1,500 s: growth of the primary species manifesting itself at $q > 1$ nm$^{-1}$ by the shift of the scattering curves towards lower $q$ values and the gradual decrease in intensity towards $I(q) \propto q^{-4}$.

To interpret the evolution of the system reflected through the observed changes in our scattering patterns during these four stages, we developed a coherent mathematical model, which

**Table 1 | Physicochemical information.**

| $[CaSO_4]$ (mmol l$^{-1}$) | [NaCl] (mmol l$^{-1}$) | $T$ (°C) | $SI_{gypsum}$ | $SI_{bassanite}$ | $SI_{anhydrite}$ |
|---|---|---|---|---|---|
| 50 | 100 | 12 | 0.50 | −0.45 | 0.20 |
| 50 | 100 | 21 | 0.50 | −0.37 | 0.28 |
| 50 | 100 | 30 | 0.49 | −0.38 | 0.36 |
| 50 | 100 | 40 | 0.49 | −0.19 | 0.45 |
| 75 | 150 | 21 | 0.70 | −0.16 | 0.48 |
| 100 | 200 | 21 | 0.63 | −0.02 | 0.84 |
| 150 | 300 | 21 | 1.04 | 0.18 | 0.83 |

Final mixed solution concentrations, salinities, temperature and saturation indices (SI) for the three $CaSO_4 \cdot xH_2O$ polymorphs.

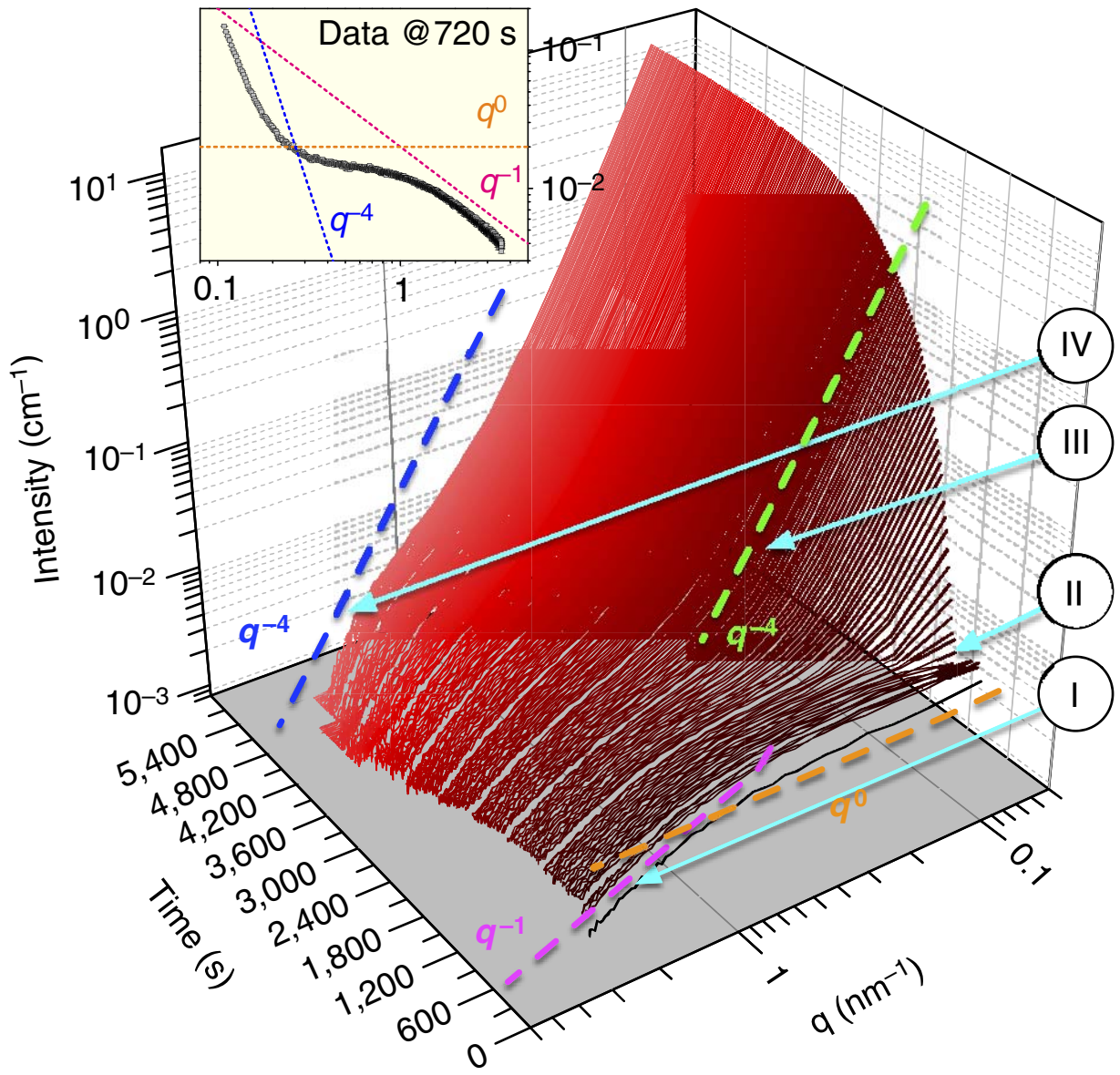

**Figure 1 | Time-resolved *in situ* SAXS patterns.** The formation of solids in the experiment with an initial concentration of 50 mmol $l^{-1}$ $CaSO_4$ and equilibrated at 12 °C; (I): formation of small primary entities/scatterers as evidenced through the change in $I(q) \propto q^{-1}$ for $q > 1\,nm^{-1}$ and $I(q) \propto q^0$ for $q < 1\,nm^{-1}$ (pink and orange dashed lines); (II) development of a structure factor; (III) formation and development of large scattering features evidenced through the increase in intensity at $q < 1\,nm^{-1}$; this change followed a dependence of $I(q) \propto q^{-3>a>-4}$ (where $a$ is the exponent—green dashed line); (IV) growth of the primary species manifesting itself at $q > 1\,nm^{-1}$ by the shift of the scattering curves towards lower $q$ values and the gradual decrease in intensity towards $I(q) \propto q^{-4}$ (blue dashed line); the inset shows a selected scattering curve and indicates the significance of the $I(q)$ dependence of the scattering exponents $q$ (pink, orange and blue dashed lines) pointing out the characteristic features in scattering as described in the main figure.

allowed us to fit the scattering curves and extract relevant information about all aspects of the scattering entities present in the reacting solutions throughout the whole length of the reaction. Only the most relevant aspects of the SAXS data analysis are included in the main text, while full details concerning the developed model are described in the Supplementary Information.

**Primary species in stages I–III**. Stages I–III (from 0 to 1,500 s) of the process involved the formation of the primary species, the interparticle interactions between and aggregation of these well-defined primary species (Fig. 2).

The first four scattering curves in Fig. 2a corresponded to stage I (30–120 s) and their shape reveal that the scattering originated from non-aggregated and non-interacting individual species of elongated, anisotropic shapes. Therefore, for stage I, the scattering curves could be best fitted with equation (1), which includes the analytical expression for a cylindrical form factor[19] $P_{cyl}(q,R,L;$ Supplementary Note 2, Supplementary Equation (1) and Supplementary Fig. 2):

$$I(q,R,L,\phi) = \phi V_{part}(\Delta\rho)^2 \cdot P_{cyl}(q,R,L) \cdot S_{eff}(q) \qquad (1)$$

where $L$ is the length, $R$ the radius, $\phi$ the volume fraction of primary scatterers, $\Delta\rho$ is the scattering length density difference between the scatterers and the matrix (solvent) and $V_{part}$ is the volume of a single particle (scatterer). $S_{eff}(q)$ represents the general expression for the effective interparticle structure factor[19]: in stage I primary species were non-interacting, hence $S_{eff}(q) = 1$, but for stages II–IV $S_{eff}(q) \neq 1$ as it is indicated by the changes at $q < 1\,nm^{-1}$ (Fig. 2a,b). The expression for the cylindrical form factor is used here as the best approximation, and we do not imply that species are rigid. In Fig. 3a, the evolution of the radius and length ($R$ and $L$) of the formed primary species, that are characterized by a cylindrical form factor, is shown up to ~1,500 s (stages I–III). Between 30 and ~800 s, the length of the formed entities, $L$, remained constant with a value of ~2.8 nm. Subsequently, these primary species gradually grew in length reaching ~4 nm at ~1,500 s (~30% increase). Their average radius was <0.3 nm and remained constant up to ~1,500 s. The fitting of the scattering curves in all other experiments yielded $R = 0.2 \pm 10\%$ nm and $L = 2.7 \pm 12\%$ nm, confirming that independent of reaction conditions, these individual primary species constituted the building blocks for the larger aggregates (Supplementary Fig. 1 and Supplementary Note 1).

The above mentioned ~30% increase in length of the primary species was also corroborated through the change in the pre-factor of the cylindrical form factor ($\phi V_{part}(\Delta\rho)^2$, equation (1)). At the earliest stage of the reaction (Fig. 3b) a very sharp increase between 30 and 60 s is followed by a constant $\phi V_{part}(\Delta\rho)^2$ value between 60 and ~800 s, followed by a gradual increase up to ~1,500 s. The initial sharp increase originates from the formation of the primary species during the solution injection/mixing period (~15 s). Between 60 and ~1,500 s, the scatterer volume normalized pre-factor $\phi(\Delta\rho)^2$ remained constant at $\sim 3.4 \times 10^{19}\,cm^{-4}$ (solid horizontal line in Fig. 3b), indicating that the primary scatterers started forming in the first 30 s, and that after 60 s no further changes were observed until 1,500 s. Since, the species at 30 s had the same $R$ and $L$ as in the consequent frames (Fig. 3a), we inferred that their electron density remained constant as well. This means that the frame at 30 s corresponded to the moment of time when the volume fraction of species, $\phi$, was temporarily lower.

**Domains of the primary species in stage II**. In stage II, for all scattering profiles shown in Fig. 2a, at $q > 1\,nm^{-1}$, neither the shape of the profiles nor the intensity values of the scattering features changed, but at $q < 0.3\,nm^{-1}$, the intensity systematically decreased between 150 and 390 s. We attributed this decrease to an evolution of the interparticle structure factor ($S_{eff}(q) \neq 1$, equation (1)), which resulted from the increase in particle–particle interferences between the primary species in solution. Such decrease in scattering intensity could be modelled by a polydisperse structure factor[20], $<S_{HS}(q)>$, that accounts for the interactions through the hard-sphere repulsive potential[21] (Supplementary Notes 3 and 4, Supplementary Fig. 3, Supplementary Equations (2–6)). In this model $<S_{HS}(q)>$ depends on three parameters: $\nu$, $<R_{eHS}>$ and $\sigma$ (Supplementary Equations (4–6)). $\nu$ represents the local volume fraction of interacting neighbouring scatterers, and as such is a measure of the local order. $<R_{eHS}>$ is an effective average hard-sphere radius, which is coupled with the associated s.d. $\sigma$. The hard-sphere radius expresses the average separation distance between the primary species equal to $2<R_{eHS}>$. $<R_{eHS}>$ varied between ~8 and ~11 nm, while the s.d., $\sigma$, showed a positive trend from an initial $\sigma/<R_{eHS}>$ value of ~7% increasing to ~40% (Fig. 3c). This indicates an increase in variance of the interparticle correlation. The evolution of the local volume fraction parameter, $\nu$, showed an increase from ~1% at 150 s to ~4.5% at 360 s, followed by a decrease to 2.5% at 510 s (Fig. 3d). Based on our analysis (Supplementary Notes 3 and 4), we propose that the primary species formed domains of locally increased scatterer number densities separated by regions depleted of scatterers, that is, local species number density fluctuations were present in the solution. The actual physical dimensions of interacting elongated primary species can be compared with the typical distance between these neighbouring entities, by relating the effective hard-sphere radii, $<R_{\mathrm{eHS}}>$ with the radii of gyration ($R_g$) of the primary species (Supplementary Note 3, Supplementary Equation (3)). The trends in Fig. 3c,d suggest that interacting domains of primary species contain individual entities separated from each other by on average $2<R_{\mathrm{eHS}}>$, and hence being un-aggregated, as $R_g$ is substantially smaller than $<R_{\mathrm{eHS}}>$. Because, the local volume fraction parameter $\nu$ expresses the degree of correlation within the particle domains[22–24], larger values indicate denser and more extensive domains, reaching a maximum at 360 s.

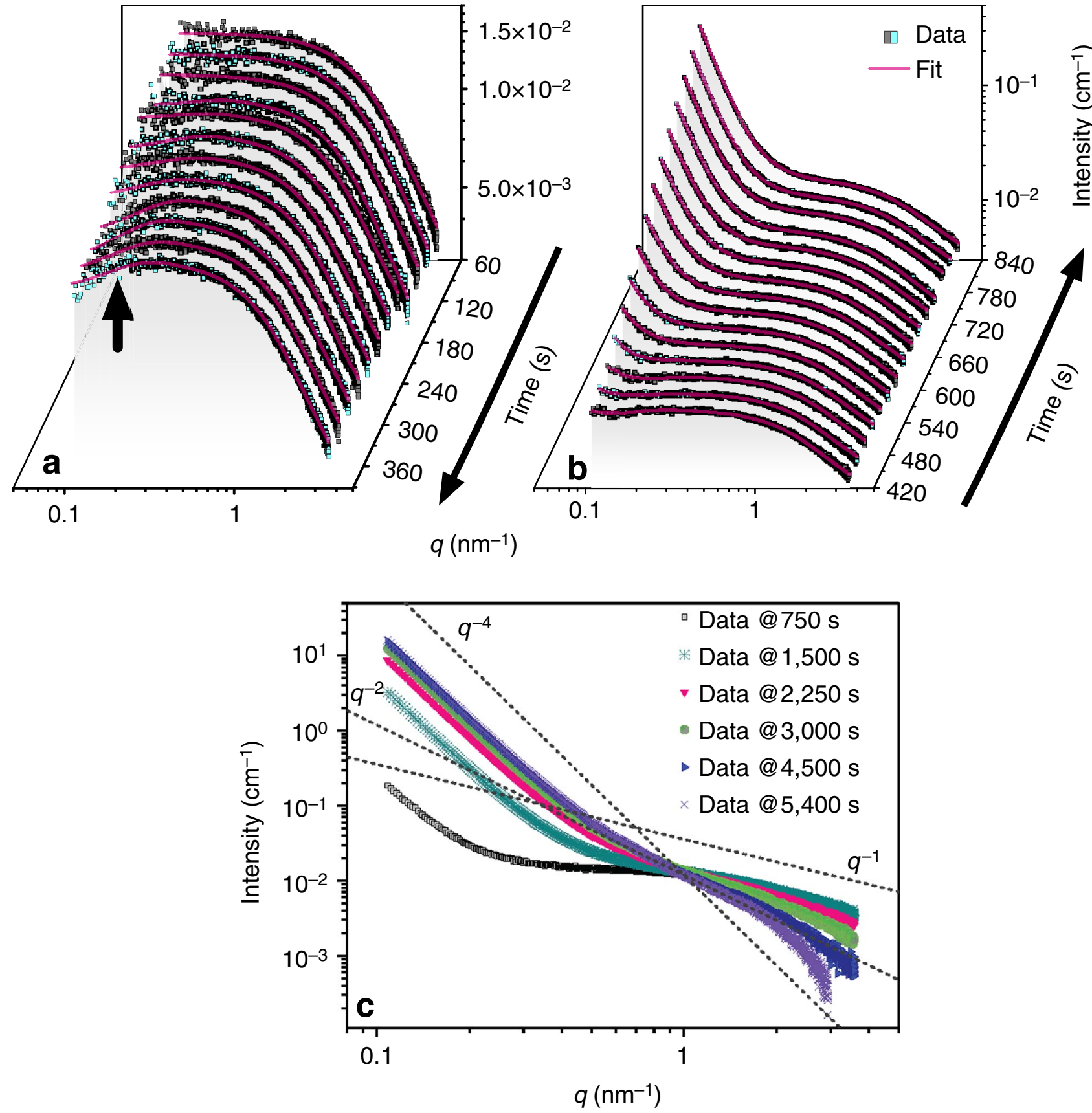

**Figure 2 | SAXS patterns in stages I–III.** (**a**) The first 390 s (stages I and II, excluding a frame at 30 s). The arrow points to the part of the curve affected by the structure factor; (**b**) Stage III as represented by the SAXS patterns between 420 and 840 s (solid lines represent best fits as described in the main text; please note that the directions of the time-axes in **a** and **b** are different); (**c**) progressive change in the intensities in the SAXS patterns between 750 and 5,400 s indicating the transitions from stages I–IV, and showing the $I(q)$ dependencies and the change in scattering exponents (dashed lines).

**The aggregation in stage III and the growth in stage IV**. During stage III between 420 and 1,500 s, in the $q>1\,\mathrm{nm}^{-1}$ region only negligible variations in shape and intensity were observed (Fig. 2b,c). On the other hand at $q<1\,\mathrm{nm}^{-1}$ a characteristic increase in intensity occurred, indicating a gradual growth of the larger scattering features. Beyond 420 s, the overall increase in scattering at $q<1\,\mathrm{nm}^{-1}$ followed a $I(q)\propto q^{-3>a>-4}$ dependence indicating that the signal could be attributed to scattering from rough fractal surfaces[25,26] (Fig. 1). We modelled the contribution of the surface fractals to scattering by introducing the expression for the structure factor $S_{\mathrm{SF}}(q)$ (Supplementary Notes 5–7, Supplementary Figs 4 and 5, Supplementary Equations (7–9)). From the contribution of $S_{\mathrm{SF}}(q)$ to the fitting, we evaluated two independent parameters characterizing the structure factor: $A'$ and $D_{\mathrm{s}}$ (see Fig. 3e). $A'$ is proportional to the relative surface (not the actual surface) of all large scattering objects formed from the aggregating primary species, because it is normalized against the form factor $P_{\mathrm{cyl}}(q)$ and the pre-factor $\phi V_{\mathrm{part}}(\Delta\rho)^2$ to fulfil the condition that $S_{\mathrm{SF}}(q\rightarrow\infty)=1$. Thus, the parameter $A$' is proportional to the number of primary scatterers in a given volume (and corresponding surface areas) at the surface of the aggregate (that is, relative surface area—see Supplementary Notes 5 and 6, Supplementary Equations (7 and 8)). The parameter $D_{\mathrm{s}}$ is a surface fractal dimension, which relates $I(q)\propto q^{-6+D_{\mathrm{s}}}$, where $D_{\mathrm{s}}=2$ represents smooth surfaces and $D_{\mathrm{s}}\rightarrow 3$ represents very rough fractal surfaces.

The surface fractal surfaces continued to develop throughout stages III and IV as indicated by the continuous increase in intensity at $q>1\,\mathrm{nm}^{-1}$ (Figs 1 and 2c) and the increase of the associated $A'$ (Fig. 3e). The onset of stage IV was, on the other hand, marked by the change in the scattering patterns at $q>1\,\mathrm{nm}^{-1}$ indicating the growth of the primary species after 1,500 s (Figs 1 and 2c). The expression for the cylindrical form factor used before had been thus no longer valid and the scattering curves between 1,500 and 5,400 s (Fig. 1) were fitted with an expression for the scattering intensity including surface fractal contribution $S_{\mathrm{SF}}(q)$ and scaled by $\phi V_{\mathrm{part}}(\Delta\rho)^2$ (equation (1); Supplementary Note 7, Supplementary Fig. 5). In the high-$q$-range, between 1,500 and 3,000 s (Fig. 2c), the scattering curves showed characteristic 'curling down', and with increasing time the scattering intensity followed a $I(q)\propto q^{-1>a>-4}$ dependence with the exponent gradually approaching a value of $-4$ at 5,400 s (Fig. 2c). The transition from stage III to IV, is reflected by the evolution of $D_{\mathrm{s}}$ with time (Fig. 3e), which shows roughening of the scattering surfaces. It is indicated by the slow increase in the fractal dimension from a temporary plateau value of $\sim$2.10 (present between 900–1,500 s), up to a value of ~2.13 at 5,400 s. The growth of the fractal surfaces in stage IV (between 1,500 and 5,400 s) is related both to the aggregation of the primary species and their transformations during stage IV. This is corroborated by the fact that after ~3,500 s the $A'$ value decreased, which is equivalent to a decrease in the total area of the scattering features. This indicates that individual scatterers coalescence and grow during stage IV, after 1,500 s.

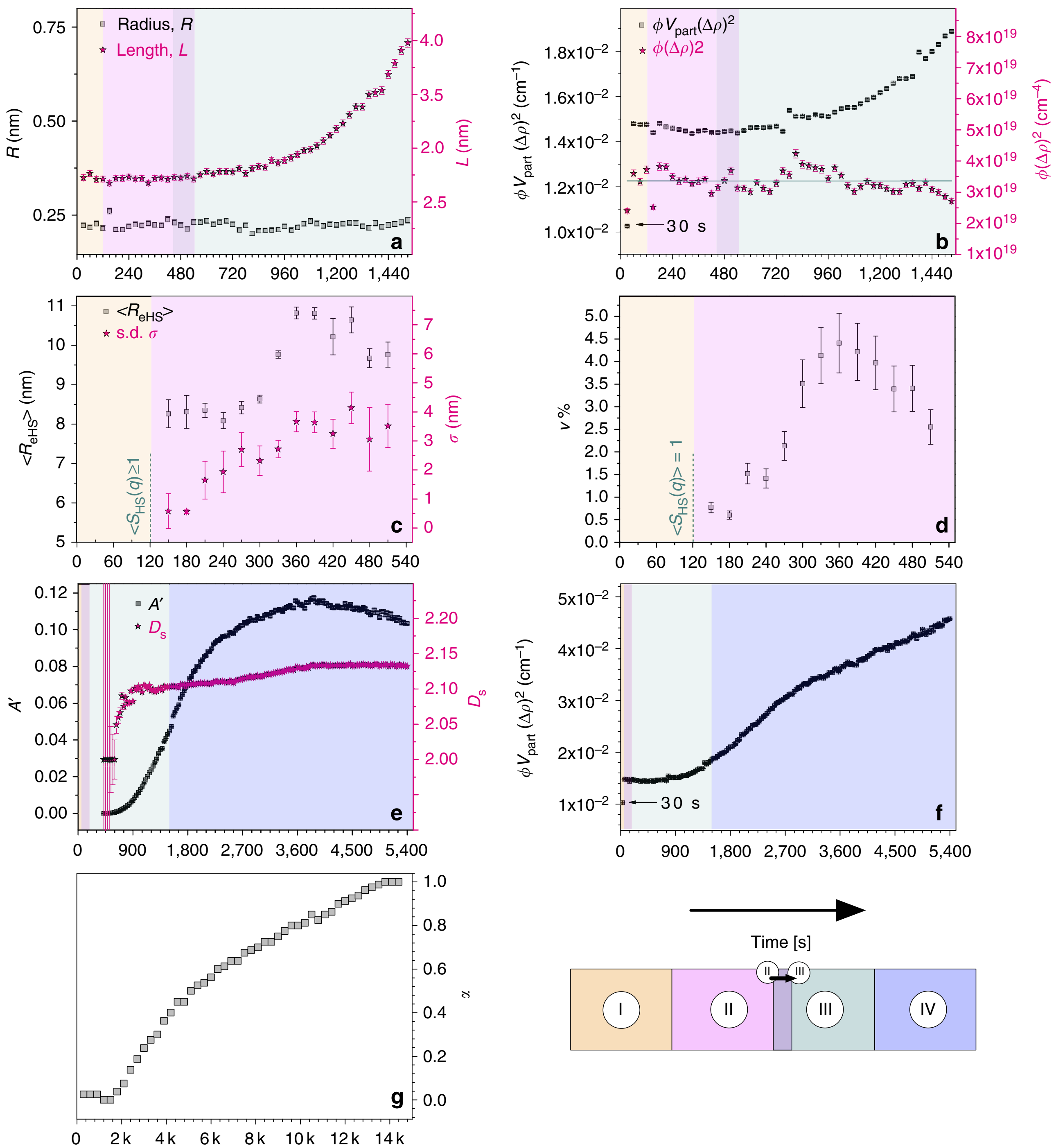

**Figure 3 | Evolution of the fitting parameters as a function of time.** (**a**) Elongated scatterers with lengths, $L$, and cross-sectional radii, $R$, up to ~1,500 s; (**b**) pre-factors $\phi V_{part}(\Delta\rho)^2$ and $\phi(\Delta\rho)^2$ up to ~1,500 s; the average of the $\phi(\Delta\rho)^2$ product is marked with a horizontal solid line; (**c**) mean effective hard-sphere radii $<R_{eHS}>$ and their s.d.'s, $\sigma$ characterizing $<S_{HS}(q)>$; (**d**) local volume fractions, $v$, characterizing $<S_{HS}(q)>$; (**e**) surface area contribution, $A'$, and surface fractal dimension, $D_s$, characterizing $<S_{SF}(q)>$. Note that up to 600 s the $A'$ and $D_s$ values exhibited very large uncertainty, due to the limited contribution of large scatterers in the low-$q$ range of the scattering patterns. For $D_s$ (<600 s) any value between 2 and <3 would produce reasonable fits, but the yielded $A'$ values for all $D_s > 2$ were significantly out of trend and are thus not shown; (**f**) pre-factor $\phi V_{part}(\Delta\rho)^2$ evolution up to 5,400 s; (**g**) Degree of crystallization α versus time from the corresponding WAXS measurements derived from the change in the area of the (020) WAXS reflection of gypsum measured simultaneously with the corresponding SAXS patterns. The background colours in **a**–**f** indicate stages I–IV, according to the legend. Time scales are in seconds (horizontal axes) except for **g** where they are in thousands of seconds.

In addition, during the latter stages of the reaction, between 1,500 and 5,400 s, the pre-factor value $\phi V_{part}(\Delta\rho)^2$ gradually increased by another ~70%, which we attribute primarily to the increase of the average volume of individual particles in stage IV (Fig. 3f). Interestingly the pre-factor values after 1,500 s did not yet reach a plateau, which suggests that rearrangement and growth of particles in stage IV was not completed at 5,400 s. Importantly, the time point of 1,500 s also corresponds with the emergence of the first diffraction peaks for gypsum in the WAXS signal (Fig. 3g). The period preceding the appearance of diffraction signals in WAXS patterns is referred to as the induction time and indicates the onset of the formation of a new crystalline phase, as shown for other systems[27,28]. In our experiments this coincides with the onset of the formation of gypsum. The crystallization of gypsum is directly evidenced through the change in area of the (020) reflection of gypsum (plotted as the parameter $\alpha$ as a function of time in Fig. 3g). This was also reflected in the changes described above for the SAXS patterns, which showed the growth of the primary species during stage IV (Fig. 3f). Furthermore from Fig. 3g it is apparent that the crystallization continued after 5,400 s (plateau not reached, Fig. 3g). At this late stage the changes in SAXS reflected the oriented rearrangement of scatterers, which was also evidenced by the development of the anisotropic scattering patterns (for details see Supplementary Note 8 and Supplementary Fig. 6).

## Discussion

The analysis of our time-resolved and *in situ* scattering data revealed that the formation of a simple salt, like gypsum, proceeds through a complex four-stage process. The first entities observable with SAXS in a $CaSO_4$ solution supersaturated with respect to anhydrite and gypsum are well-defined, sub-3 nm in length, species. The initial constant values of the $\phi V_{part}(\Delta\rho)^2$ and the normalized $\phi(\Delta\rho)^2$ pre-factors (Fig. 3b) suggest that: (i) the primary species formed near-instantaneously after mixing of the $Ca^{2+}$- and $SO_4^{2-}$-stock solutions (that is, stage I, within 30 s); (ii) although the length of these primary species grew by ~30%, their volume fraction remained unchanged throughout stages I–III, and their number density decreased. Hence, the increase in length occurred by the merging of primary units to form longer ones. Also noteworthy are the density fluctuations occurring in stage II, which indicate that the primary species form a mixture of denser (of increasing local volume fraction Fig. 3d) and less-dense domains in the bulk of the solution. In recent years, the pivotal role of density fluctuations in the nucleation process has been postulated[29,30] and observed in various simulation studies of hard-spheres[31].

Inexorably the question arises about the nature of these primary species and their evolution through time. Monitoring the $\phi V_{part}(\Delta\rho)^2$, and the normalized $\phi(\Delta\rho)^2$ pre-factors (Fig. 3b,f), allowed us to analyse the reaction progress as a function of time. However, specific information concerning the electron density, $\Delta\rho$, and volume fraction, $\phi$, of the forming phase is not independently accessible because the pre-factors are the product of these two components. Nevertheless, the scattered intensity is expressed in absolute units, and thus the changes in the $\phi(\Delta\rho)^2$ pre-factor in relation to known $CaSO_4$ polymorphs and their expected volume fractions can be estimated. This way we can correlate our scattering data with thermodynamic solubility data, and identify the formed phase(s) based on their electron densities. We evaluated the predicted volume fractions, $\phi$, for each possible $CaSO_4$ phase by considering the original concentration of $Ca^{2+}$ and $SO_4^{2-}$ ions in solution and using the various bulk solubility of each phase calculated with PHREEQC (ref. 32). Using the values of $\phi$ , we calculated the corresponding electron densities from our SAXS data (pseudo-phases, Supplementary Note 9 and Supplementary Table 2), and compared them to the actual electron densities of each of the three $CaSO_4$ phases (Fig. 4).

An unequivocal identification of the nature of the primary species through their electron density is hindered by the fact that the actual solubilities of these nano-sized primary species are not known. It is likely that at the early stages of the reaction, when the nano-sized primary units form, the process is out of equilibrium and their solubilities would be considerably different from their bulk counterpart (for example, the Ostwald-Freundlich relation[33,34]). In any case, our considerations refer to the full range of possible (nano- to bulk-) solubility values in the $CaSO_4$ system and for all conditions the calculated volume fractions yield relatively high-electron density ranges for each of the crystalline $CaSO_4$ polymorphs (Fig. 4).

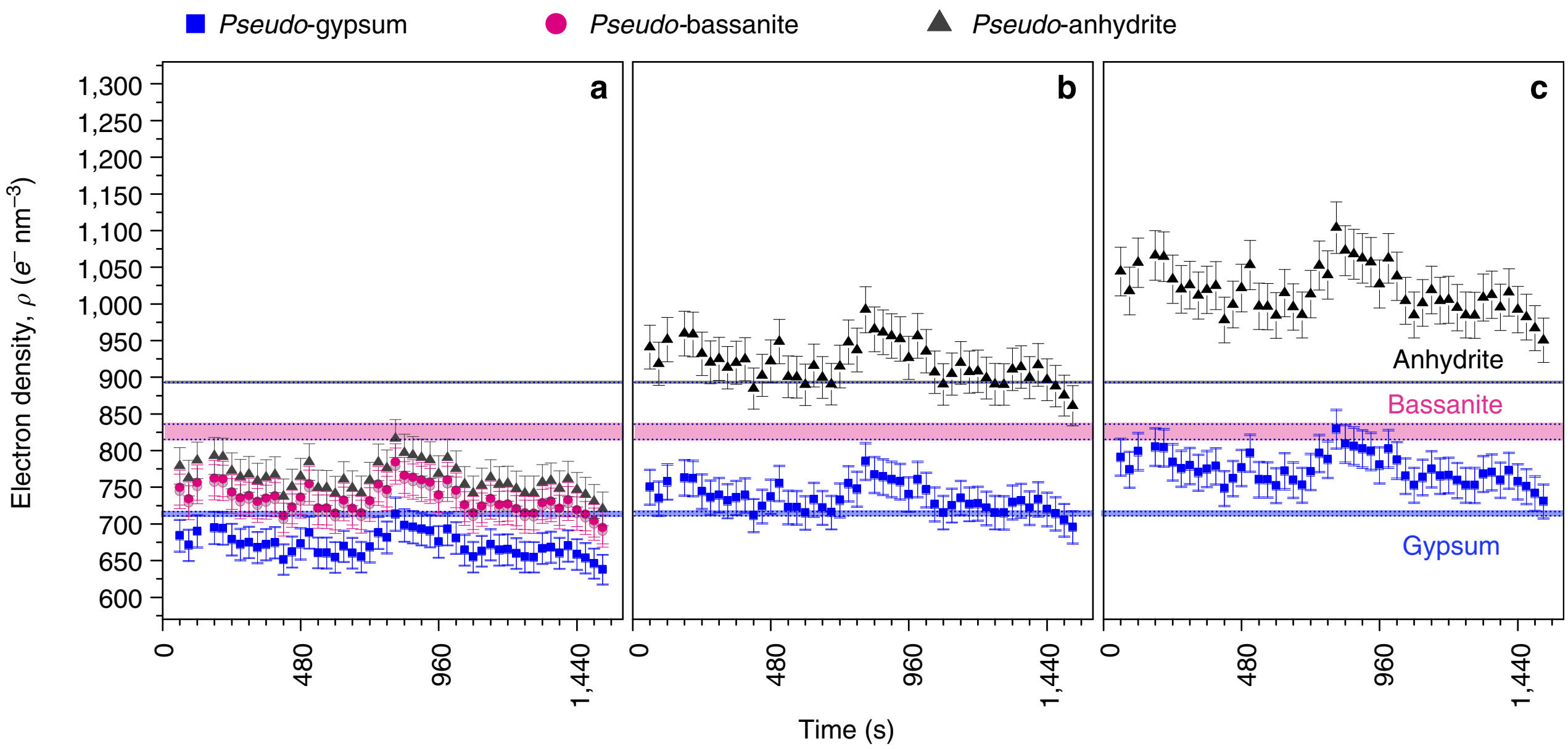

**Figure 4 | Electron densities for $CaSO_4$ pseudo-phases based on SAXS.** The values as a function of time were calculated based on the corresponding volume fractions of the known crystalline phases: (**a**) not taking into account the bulk solubility; (**b**) taking into account bulk solubility in pure water; (**c**) taking into account bulk solubility in a 100-mmol l$^{-1}$ NaCl solution. The horizontal bands indicate the range of electron densities expected for each given $CaSO_4$ polymorph. Details of how the electron density values were calculated can be found in Supplementary Note 9 and Supplementary Table 2.

Important to note is the fact that this observation is true even for the unlikely case where the bulk solubility was assumed to be zero (Fig. 4a). We consider, the zero solubility case because it determines the highest physically possible yield of the reaction, and hence the highest possible volume fraction of the primary species. Since, $\phi(\Delta\rho)^2 = \text{const}$, it also determines the lowest possible electron density of these species. The obtained values for this scenario indicate that the electron density of the observed primary species is comparable to that of gypsum. Although the exact solubility of the primary species is unknown, it is highly unlikely to be zero. This implies that their electron densities must be either equal or in fact higher than that of gypsum because, with increasing solubility, the volume fraction decreases, and in turn the electron density increases and reaches values of bassanite or anhydrite (Fig. 4b,c). Therefore, the primary species could be structurally any of the three $CaSO_4$ phases, but due to the lack of quantitative information about the solubilities of these phases at the nanoscale, it is impossible to pinpoint which one. Regardless, their electron density is very high and thus internally they cannot be hydrated more than gypsum. This does not exclude the possibility that these primary species can be hydrated at their surfaces. However, at this length-scale, this is almost impossible to detect with SAXS due to the scattering contrast, which originates from the structural differences between the primary species and the aqueous medium. On the other hand, the electron densities of the three possible $CaSO_4$ polymorphs at the local scale of the anhydrous Ca–$SO_4$-cores (that is, excluding water) are practically the same[35–37]. In other words, the differences in the bulk electron densities of the polymorphs (Supplementary Table 2) originate only from the presence of water in the structures. In Supplementary Fig. 7, we present the cross-sections of Ca–$SO_4$-cores based on the three polymorphs (anhydrite, bassanite and gypsum). For each case the cross-sections correspond to the dimensions of the primary scatterers: ~2.8-nm-long units of a radius of ~0.5 nm. We do not have any evidence as to the exact structures of the observed primary species, but with this figure we want to emphasize that at the local scale of anhydrous Ca–$SO_4$ cores, all three polymorphs appear structurally similar. Characteristically, due to the apparent geometrical restrictions of these structures, they are highly unlikely to contain any water internally. Therefore, we propose that the formation of the $CaSO_4$ solids in our experiments proceed via the aggregation of such anhydrous Ca–$SO_4$-cores that latter rearrange, within the aggregates, to form the structure of gypsum, bassanite or anhydrite. This is further supported when one considers in detail the processes occurring in stage IV.

Finally, the question arises whether these proposed Ca–$SO_4$-cores should be considered pre-nucleation clusters (PNC), as previously reported for the $CaCO_3$ system by Gebauer *et al.*[38,39]. PNCs have been inferred from titrations and ultracentrifugation data sets and this was done starting from solutions undersaturated with respect to all bulk $CaCO_3$ phases. So far the presence of PNCs in undersaturated solutions has not been confirmed through SAXS measurements, likely due to the low signal-to-noise ratios. Our $CaSO_4$ SAXS data were however collected only at supersaturated conditions with respect to gypsum, and thus no meaningful conclusions about the existence of PNCs can be made. Furthermore, PNCs are usually considered to be dissolved species and thus they do not constitute a separate phase and are not solid particles. X-ray scattering is sensitive to any fluctuations in the local electron density at the length scales within the $q$-range measured and this is regardless of the thermodynamic nature or species in solution. At the same time the actual thermodynamic categorization of species is not often reflected through their structure. We think that the true nature of PNCs can only be inferred from structural investigations (both scattering and imaging) probing a broad spectrum of conditions, which has not been done for any mineral system.

The actual crystallization of gypsum corresponds to the coalescence and growth in all directions of the primary scatterers within the surface fractal aggregates. This occurs only in stage IV, while stages I–III coincide with the induction period observed in WAXS (Fig. 3g and schematics in Fig. 5). Importantly, the arrangement of the primary species in the aggregates formed in stage III does not contribute to a coherent diffraction signal despite the large size of those aggregates. The very fact that primary species are observable in scattering at $q > 1\,\text{nm}^{-1}$ simultaneously with the surface fractals at $q < 1\,\text{nm}^{-1}$ indicates that there must be some sort of misalignment and voids in between the primary species within the aggregates (that is, disordered brick wall structure, Fig. 5c). Most likely these voids are filled with water and unreacted ions, which take part in the rearrangement, coalescence and crystallization processes of stage IV, and involve also the organization of water molecules into the gypsum structure. Typically, for particles aggregating in solution one would expect mass-fractal-like aggregates, because such aggregates form at relatively low-local particle concentrations where the diffusion length is considerable[40–42]. However, Kolb and Herrmann[43] showed through Monte Carlo simulations of highly concentrated colloidal aggregates, that if the 'local' concentration is close to 1, than surface fractal aggregates are formed instead of their mass (volume) counterparts. In our case such surface fractal aggregates likely formed due to the collapse of the primary species-rich domains from stage II. In stage II, when the primary species within the domain were still separated from each other by a distance of $2\langle R_{eHS}\rangle$ (Fig. 3c) the local volume fraction $\nu$ was roughly 10–20 times higher than the actual (that is, global) volume fraction $\phi$ (compare Supplementary Note 9 and Fig. 3d). A collapse of such a domain structure would lead to a high-local concentration of primary species. Nevertheless, such a process does not imply that sufficient structural coherence, to the extent necessary for diffraction, could occur within the aggregate. Only on further transformations (growth and re-organization of primary species), forming better-ordered coherent structures (Fig. 5d), can a diffraction pattern arise.

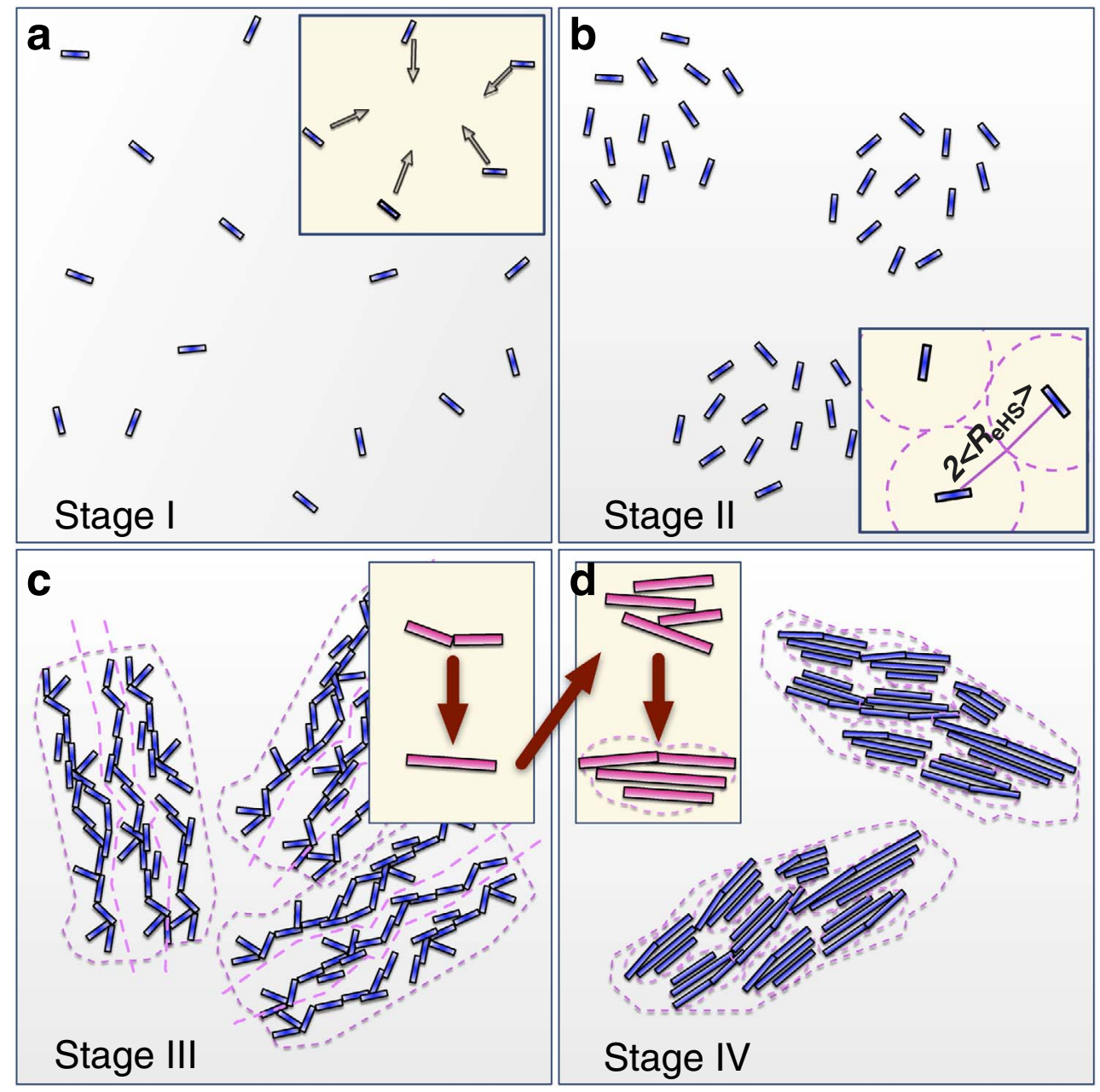

**Figure 5 | Schematic representation of the four stages of $CaSO_4$ formation.** (**a**,**b**) As observed for our experimental conditions: (stage I) formation of well-defined, sub-3 nm primary species/scatterers; (stage II) formation of domains of primary species; inset shows that scatterers are on average separated by a distance of $2\langle R_{eHS}\rangle$; (stage III) aggregation and self-assembly of the primary species forming large surface fractal morphologies; (stage IV) growth and coalescence of the primary species within the aggregates; insets in **c** and **d** show the consequent stages of growth—increase in length followed by increase in all dimensions eventually leading to larger morphologies.

Recent experimental results show that at room temperature the outcome of the $CaSO_4$ formation reaction in terms of phase selection can be directed by either changing the available amount of water; that is, by using different mixtures of water—ethanol or the salinity[44–46]. These studies showed that phase formation in the $CaSO_4$ system can be guided to form gypsum, bassanite, anhydrite or a mixture of these phases. Based on these experimental observations, and those presented in this current work, we propose that the aggregates formed in stage III have a framework structure, made of anhydrous Ca–$SO_4$-core-based primary species and disordered water, which is common to all $CaSO_4$ phases (Supplementary Fig. 7). In stage IV to obtain well-ordered sheets of $CaSO_4$-cores and $H_2O$ layers (as found in gypsum), the primary species within the aggregates from stage III, must radically transform and coalesce to larger particles in stage IV. Thus, the most likely path towards the final gypsum polymorph proceeds through an internal rearrangement of the aggregates, assuming a local less-ordered structure with respect to water, in the early part of stage IV. Subsequently this less-ordered phase transforms to the dihydrated gypsum in which the water structuring is more complex, making bassanite a plausible nano-phase on the pathway to gypsum formation if the reaction is quenched[12–14,16]. Consequently, if not enough water is available the reaction will be directly halted at this intermediate point, and bassanite will emerge as the bulk crystalline phase, as was previously reported[44–46].

The precipitation pathway described in this work could well be the key to explain the abundant presence and long-term persistence of bassanite on Mars (compared with very rare presence on Earth). The surface conditions of Mars are characterized by extreme low-water activities both now and in the past[47]. Considering the presence of current Mars sol-night-time transient liquid brines[48], may help infer that on Mars both the formation of $CaSO_4$ phases from supersaturated (even brine-like) solutions may follow the four-stage process described above (Fig. 5) but that the stabilization of a thermodynamically unstable phase-like bassanite is possible on Mars due to the low availability of water (which if present would, through continual hydration lead to the transformation of bassanite to gypsum).

On the other hand, the concept of a framework structure, common to all three crystalline $CaSO_4$ phases, represents a novel way of looking at the poorly understood formation mechanism(s) of different polymorphs for a given system, and could open the door for new ways of controlling polymorph selection. In addition, with the current knowledge of the precipitation pathway new strategies to design more effective anti-scalants or more energy efficient ways to stabilize bassanite for construction purposes can be envisaged. Our data suggest that targeting stages II and III of our pathway may pave the path to the production of stable bassanite through more efficient means than dehydration upon heat-treatment.

## Methods

**Synthesis of calcium sulfate phases.** $CaSO_4$ was synthesized by reacting equimolar aqueous solutions of $CaCl_2 \cdot 2H_2O$ (pure, Sigma) and $Na_2SO_4$ (>99%, Sigma) at final concentrations of the mixed solutions of 50, 75, 100, 150 $\text{mmol l}^{-1}$ at $T = 21\,°C$, and also for 50 $\text{mmol l}^{-1}$ at $T = 12$, 30, 40 °C. Before mixing, all solutions were equilibrated at the desired reaction temperatures and filtered through 0.2 μm pore size polycarbonate filters to remove possible impurities. The saturation indices (SI, Table 1) with respect to the different bulk calcium sulfate hydrates and at the salinities/ionic strengths (including contributions for the $Na^+$ and $Cl^-$ counter-ions) were calculated with PHREEQC (ref. 32) based on the following reaction:

$$CaCl_2 \cdot 2H_2O + Na_2SO_4 \rightarrow CaSO_4 \cdot xH_2O \downarrow + 2NaCl$$

Only bulk solubility data are available for all three polymorphs. The SI values show that under the indicated physicochemical conditions all mixed solutions were supersaturated with respect to gypsum and anhydrite ($SI_{gypsum} > 0$, $SI_{anhydrite} > 0$), whereas bassanite was in all cases, except for 150 $\text{mmol l}^{-1}$ ($SI_{bassanite} > 0$), undersaturated. Thus, bassanite should not precipitate but dissolve ($SI_{bassanite} < 0$).

***In situ* set-up description.** All $CaSO_4$ formation reactions were performed in a 200-ml temperature-stabilized glass reactor. The reacting solutions were continuously stirred at 350 r.p.m., and circulated through a custom-built PEEK flow-through cell with embedded quartz capillary (external diameter 1.5 mm, wall thickness ~10 μm) using a peristaltic pump (Gilson MiniPuls 3, flow ~10 ml per s). Typically an experiment started with 40 ml of a temperature-stabilized $CaCl_2$ aqueous solution inside the reactor. This solution was circulated through the capillary cell while scattering patterns were collected continuously as described below. $CaSO_4$ formation reactions were initiated remotely through the injection of 40 ml of a temperature-stabilized $Na_2SO_4$ aqueous solution. Fast injection and mixing (within 15 s) of the two solutions was achieved with the use of the fast-injection mode of a stopped-flow system (Bio-Logic SFM-400). Depending on reaction conditions (supersaturation or temperature) reactions were followed up to 4 h.

**Characterization with scattering methods.** All *in situ* and time-resolved small- and wide-angle X-ray scattering (SAXS/WAXS) measurements were carried out at beamline I22 of the Diamond Light Source Ltd (UK). Experiments were performed using a monochromatic X-ray beam at 12.4 keV and two-dimensional (2D) scattered intensities were collected at small-angles with a Dectris Pilatus 2 M (2D large area pixel-array detector[49]). Transmission was measured by means of a photodiode installed in the beam-stop of the SAXS detector. A sample-to-detector distance of 4.22 m allowed for a usable $q$-range of $0.1 < q < 3.8\,nm^{-1}$. The scattering-range at small-angles was calibrated against silver behenate[50] and dry collagen standards[51]. For the wide-angle measurements we used a HOTWAXS detector (a photo-counting one-dimensional (1D) microstrip gas chamber detector[52]). The WAXS detector was calibrated with synthetic and highly crystalline silicon (NIST SRM 640C), and with commercial gypsum and bassanite powders (Sigma Aldrich). The scattered intensity was calibrated to absolute units using water as a reference.

The reactions were followed *in situ* from the very early stages and up to the point when a final crystalline phase was fully developed and no more changes in the WAXS peak intensities were observed. Furthermore, for each experiment we also measured a series of backgrounds and reference samples including: the empty capillary cell, cell filled with water and cell filled with the initial, unmixed $CaCl_2$ and $Na_2SO_4$ solutions at the various used concentrations and temperatures. In all simultaneous SAXS/WAXS measurements, the acquisition time per frame varied between experiments (from 1 to 30 s per frame) and this time frame was based on previously off-line tested reaction times for the various conditions. The triggering of SAXS and WAXS frame acquisition was synchronized between the two detectors, so that a given frame in SAXS corresponded to the one in WAXS. Most of the recorded 2D SAXS patterns were found to be independent of the in-plane azimuthal angle with respect to the detector (that is, scattering patterns where circular in shape), showing that the investigated systems could be considered isotropic. In those patterns, pixels corresponding to similar $q$ regardless of their azimuthal angle where averaged together, and hence the 2D patterns were reduced to 1D curves. In several cases dependence between the scattering intensity and the in-plane azimuthal angle was observed (that is, the scattering patterns were elliptical in shape). This indicated preferred orientation of the scatterers in the investigated samples. Therefore, selected angle-dependent 1D scattering curves were obtained by averaging of pixels with similar $q$ and limited to ca. ± 3° angle off the direction indicated by the chosen azimuthal angle: the equatorial and meridional directions of the elliptical 2D patterns. SAXS data processing and reduction included primarily masking of undesired pixels, normalizations and correction for transmission, background subtraction and data integration to 1D. These steps were performed using the Data Analysis WorkbeNch (DAWN) software package (v. 1.3 and 1.4) according to I22 guidelines[53].

For WAXS data, to increase the signal-to-noise ratio, the collected diffraction patterns were averaged together maintaining the same proportion of added frames for the total course of the experiment for each data set, thus allowing for more accurate characterization of the potential calcium sulfate phases present in solution. The *in situ* diffraction indicated after a long induction time only the presence of gypsum. The time-resolved WAXS patterns were fitted using XFit-Koalariet[54], which allowed us to extract the degree of crystallization, α, over the course of the formation of gypsum.

## Acknowledgements

This work was made possible by a Marie Curie grant from the European Commission in the framework of the MINSC ITN (Initial Training Research network), project number 290040. Access to the Diamond Light Source Ltd (UK) for beamtime was possible through a direct access grant to L.G.B.; Beamline staff of I22, A. J. Smith and N. J. Terrill, are thanked for their competent support and advice.

## Author contributions

The manuscript was written through equal contributions of all authors and all authors have given approval to the final version of the manuscript.

## Additional information

**Supplementary Information** accompanies this paper at http://www.nature.com/naturecommunications

**Competing financial interests:** The authors declare no competing financial interests.

**Reprints and permission** information is available online at http://npg.nature.com/reprintsandpermissions/

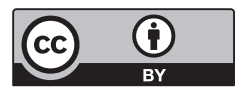